\documentclass[sigconf]{acmart}
\AtBeginDocument{%
  }

\copyrightyear{2026}
\acmYear{2026}
\setcopyright{cc}
\setcctype{by}
\acmConference[WSDM '26] {Proceedings of the Nineteenth ACM International Conference on Web Search and Data Mining}{February 22--26, 2026}{Boise, ID, USA.}
\acmBooktitle{Proceedings of the Nineteenth ACM International Conference on Web Search and Data Mining (WSDM '26), February 22--26, 2026, Boise, ID, USA}
\acmISBN{979-8-4007-2292-9/2026/02}
\acmDOI{10.1145/3773966.3777954}
\usepackage{xcolor}
\usepackage{colortbl}
\usepackage{graphicx}
\usepackage{makecell}
\usepackage{tabularx}
\usepackage{multirow}
\usepackage{xspace}
\usepackage{enumitem}
\usepackage{float}
\usepackage{longtable}
\usepackage[most]{tcolorbox}
\usepackage{colortbl}
\usepackage{adjustbox}
\usepackage{pifont}
\usepackage{balance}
\usepackage{tablefootnote}

\usepackage{etoolbox}

\newcommand{\ourname}{RecBench+\xspace}
\newcommand{\firstname}{Condition-based\xspace}
\newcommand{\secondname}{User Profile-based\xspace}

\definecolor{mycolor}{RGB}{134,150,167}
\definecolor{backred}{RGB}{255, 190, 190}
\definecolor{backblue}{RGB}{210, 230, 250}
\definecolor{mygreen}{RGB}{27,135,73}
\newcommand{\redcross}{\textcolor{red}{\ding{55}}}  
\newcommand{\greentick}{\textcolor{mygreen}{\ding{51}}} 
\usepackage{color}

\begin{document}

\newtcbtheorem[auto counter, number within = section]{cmt}{}{
	colbacktitle = mycolor, colframe = mycolor,
	colback = mycolor!10!white,
 % colback=Emerald!10,colframe=cyan!40!black
	fonttitle=\bfseries,
 % fontupper=\itshape,
}{t}

\title{Towards Next-Generation Recommender Systems: a Benchmark for Personalized Recommendation Assistant with LLMs}

\thanks{\dag Correspondence to: Wenqi Fan and Qing Li, Department of Computing, The Hong Kong Polytechnic University.}
\author{Jiani Huang}
\email{jianihuang01@gmail.com
}
\orcid{0000-0001-7016-982X}
\affiliation{%
  \institution{The Hong Kong Polytechnic University}
  \city{Hong Kong}
  \country{China}
}

\author{Shijie Wang}
\email{shijie.wang@connect.polyu.hk}
\affiliation{%
  \institution{The Hong Kong Polytechnic University}
  \city{Hong Kong}
  \country{China}
}

\author{Liang-bo Ning}
\email{BigLemon1123@gmail.com}
\affiliation{%
  \institution{The Hong Kong Polytechnic University}
  \city{Hong Kong}
  \country{China}
}

\author{Wenqi Fan\dag}
\email{wenqi.fan@polyu.edu.hk}
\affiliation{%
   \institution{The Hong Kong Polytechnic University}
  \city{Hong Kong}
  \country{China}
}

\author{Shuaiqiang Wang}
\email{shqiang.wang@gmail.com}
\affiliation{%
  \institution{Baidu Inc.}
  \city{Beijing}
  \country{China}
}

\author{Dawei Yin}
\email{yindawei@acm.org}
\affiliation{%
  \institution{Baidu Inc.}
  \city{Beijing}
  \country{China}
}

\author{Qing Li\dag}
\email{qing-prof.li@polyu.edu.hk}
\affiliation{%
   \institution{The Hong Kong Polytechnic University}
  \city{Hong Kong}
  \country{China}
}

\renewcommand{\shortauthors}{Jiani Huang et al.}
%% No italics, no superscripts, not anonymous
%% Use footnote or author note to identify equal contribution and/or contact author info

\begin{abstract}
  Recommender systems (RecSys) are widely used across various modern digital platforms and have garnered significant attention. 
Traditional recommender systems usually focus only on fixed and simple recommendation scenarios, making it difficult to generalize to new and unseen recommendation tasks in an interactive paradigm. 
Recently, the advancement of large language models (LLMs) has revolutionized the foundational architecture of RecSys, driving their evolution into more intelligent and interactive personalized recommendation assistants. 
However, most existing studies rely on fixed task-specific prompt templates to generate recommendations and evaluate the performance of personalized assistants, which limits the comprehensive assessments of their capabilities. 
This is because commonly used datasets lack high-quality textual user queries that reflect real-world recommendation scenarios, making them unsuitable for evaluating LLM-based personalized recommendation assistants. 
To address this gap, we introduce \ourname, a new dataset benchmark designed to assess LLMs' ability to handle intricate user recommendation needs in the era of LLMs. 
\ourname encompasses a diverse set of queries that span both hard conditions and soft preferences, with varying difficulty levels.  
We evaluated commonly used LLMs on \ourname and uncovered below findings: 
1) LLMs demonstrate preliminary abilities to act as recommendation assistants, 
2) LLMs are better at handling queries with explicitly stated conditions, while facing challenges with queries that require reasoning or contain misleading information. 
Our dataset has been released at \href{https://github.com/jiani-huang/RecBenchPlus}{https://github.com/jiani-huang/RecBenchPlus}.
\end{abstract}

\begin{CCSXML}
<ccs2012>
<concept>
<concept_id>10002951.10003260.10003261.10003271</concept_id>
<concept_desc>Information systems~Personalization</concept_desc>
<concept_significance>500</concept_significance>
</concept>
</ccs2012>
\end{CCSXML}
\ccsdesc{Information systems~Personalization}

\keywords{Recommender System, Recommendation Assistant, Large Language Models (LLMs), Dataset Benchmark.}


\setcopyright{none}
\maketitle

% \vspace{-1em}

\begin{figure*}[t]
    \vspace{-1.2em}
    \centering
    \includegraphics[width=.86\linewidth]{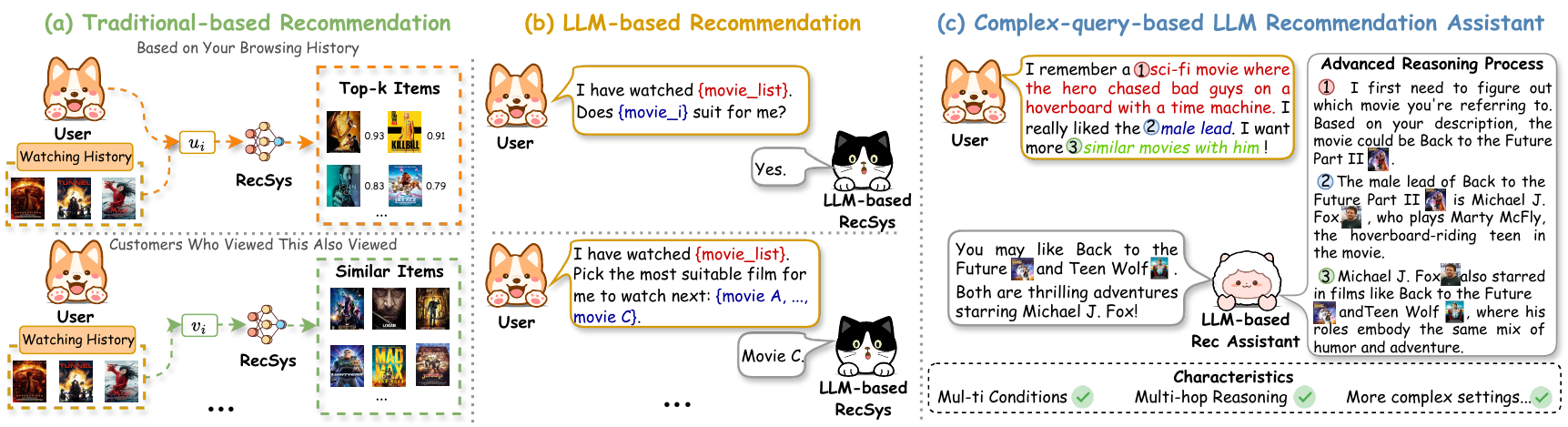}
    \vspace{-1.2em}
    \caption{\textnormal{Illustration of different recommendation paradigms.
    \textbf{(a) Traditional-based:} Generating recommendations in fixed scenarios by learning user/item representations. 
    \textbf{(b) LLM-based recommendation} Employing fixed and simple query templates, lacking evaluation in more complex scenarios. 
    \textbf{(c) Ours:} Evaluating the LLM’s ability to handle complex, flexible user queries as a recommendation assistant.}}
    \label{fig:task_illustration}
    \vspace{-1.4em}
\end{figure*}

Recommender Systems (\textbf{RecSys}) play a crucial role in modern digital platforms, enabling personalized content delivery across domains such as e-commerce~\cite{chen2023knowledge}, entertainment~\cite{gao2023chat}, and education~\cite{urdaneta2021recommendation}. 
For example, YouTube once reported that 60\% of the clicks on its homepage came from recommended content~\cite{davidson2010youtube}, revealing the huge commercial value of RecSys.
Generally, these systems are trained to handle typical recommendation scenarios, such as ``\emph{Customers Who Viewed This Also Viewed}'' to recommend alternatives or similar items 
and ``\emph{Based on Your Browsing History}'' to recommend Top-K items of interest using a user’s historical data, as shown in Figure~\ref{fig:task_illustration} (a). 
As one of the most representative algorithms, matrix factorization (MF) techniques~\cite{koren2009matrix} are proposed to learn latent representations of users and items based on user-item interactions for predicting matching scores in recommendations. 
However,  these traditional recommendation methods are applicable only to a limited range of simple and fixed recommendation scenarios, relying on pre-designed and well-trained recommendation methods, and they struggle to generalize to new and unseen recommendation tasks (i.e., recommendation queries) in an interactive paradigm.  
For instance, if a user searches for \emph{`a durable laptop for graphic design under \$1500'}, 
traditional approaches like MF depend only on the correlation between users and items, failing to capture these nuanced requirements and deliver appropriate recommendations.

Recently, large language models (\textbf{LLMs}) like GPT-o1~\cite{jaech2024openai}, DeepSeek-R1~\cite{guo2025deepseek}, and LLaMA~\cite{dubey2024llama} have demonstrated strong capabilities in generalization and reasoning. 
% \wq{For instance, in complex reasoning tasks across physics, biology and chemistry, GPT o1-mini has surpassed PhD level~\cite{jaech2024openai}. 
For instance, AUTOGLM~\cite{liu2024autoglm} acts as an autonomous assistant, leveraging the LLMs' human-like intelligence in understanding and planning capabilities to help users control digital devices via GUIs and address complex tasks such as booking flights~\cite{singh2024personal}, collecting data~\cite{schmidgall2025agent}, and coding~\cite{kazemitabaar2024codeaid,pinto2024lessons}, significantly enhancing convenience in people's daily lives. 
Inspired by the powerful capabilities of LLMs, many efforts have been made to incorporate LLMs into RecSys as personalized recommendation assistants in an interactive paradigm.  
For example, LLaRA~\cite{liao2024llara} directly transforms users' historical interactions into natural language formats (i.e., prompts) for personalized recommendations.
Benefiting from LLMs, as shown in Figure~\ref{fig:task_illustration} (b), these personalized recommendation assistants can handle more interactive and challenging scenarios, such as \textit{``Tell You What You Like,''} which allows users to ask questions in a conversational manner and receive corresponding recommendations.

Despite the remarkable success, current studies lack a comprehensive evaluation of the recommendation capability of these LLM-based personalized recommendation assistants.  
Specifically, in the phases of both training and testing, most existing methods leverage fixed and simple task-specific prompt templates to generate recommendation responses, and assess the recommendation performance of these assistants, such as  ``\emph{Will the user like \{movie\_i\}. Please answer Yes or No.}'' to determine whether to recommend a specific item, or ``\emph{Please choose a movie from the candidate list \{movie\_1, movie\_2, ... \}}'' to recommend a user-liked item. 
However, this evaluation paradigm fails to align with real-world practical scenarios, where the user queries can be highly complex when they interact with the LLM-based recommendation assistant. Consider the complex query illustrated in Figure~\ref{fig:task_illustration} (c), 
the LLM-based recommendation assistant must first interpret the user's description to identify the correct movie, then recognize the lead actor, and finally suggest similar films featuring that actor.
This process requires the assistant to understand the query, perform multi-step reasoning, and consider multiple conditions—capabilities that go beyond basic attribute matching or shallow semantic understanding. Therefore, the results of existing LLM-based RecSys on traditional benchmark datasets may not apply to the new recommendation paradigm in the LLM era. 

Existing frequently used datasets, such as LastFM~\cite{schedl2016lfm}, Amazon Beauty~\cite{ni2019justifying}, and Movielens-1M~\cite{harper2015movielens}, typically only include basic user information, item metadata (e.g., title), and user-item interactions.  
These datasets are more suitable for assessing the accuracy performance of traditional RecSys but are not appropriate for effectively evaluating recommendation assistants in the era of LLMs, due to the lack of diverse complex user queries that can meet practical recommendation scenarios. 
To bridge this gap, we introduce a new benchmark, \textbf{\ourname}, tailored for evaluating the potential of LLMs to act as personalized recommendation assistants. 
Specifically, 
we propose \textbf{a new dataset} of approximately 30,000 complex user queries, covering a range of recommendation scenarios that require complex reasoning, including multiple conditions, multi-hop reasoning, misleading information, and diverse user profiles. This dataset is built from several commonly used recommendation datasets (e.g., Movielens-1M~\cite{harper2015movielens} and Amazon book\footnote{https://archive.ics.uci.edu/ml/datasets/Amazon+book+reviews}). 

We comprehensively evaluate the performance of seven widely used LLMs on \ourname, including GPT-4o~\cite{hurst2024gpt}, Gemini-1.5-Pro~\cite{team2024gemini}, DeepSeek-R1~\cite{guo2025deepseek}, etc. 
We further analyze various factors that impact model performance, such as the number of conditions, the integration of user interaction history and different fine-tuning paradigms. Through extensive experiments with state-of-the-art LLMs in our benchmark, we obtain several insightful findings:
(1) \textbf{LLMs demonstrate preliminary abilities to act as recommendation assistant, but their strengths vary.} GPT-4o and DeepSeek-R1 excel at user queries with clear conditions, while Gemini-1.5-Pro and DeepSeek-R1 perform better in queries without clear conditions but require a clear understanding of user profiles.
(2) \textbf{Fine-tuning notably boosts LLM recommendation performance, with a two-stage approach combining supervised fine-tuning (SFT) and reinforcement fine-tuning (RFT) performing best overall.} This suggests that SFT effectively warms up the model for the task, enabling RFT to further refine performance.
(3) \textbf{For queries with user-specified conditions, LLMs are more adept at handling those with explicitly stated conditions.} 
In contrast, queries that require reasoning or contain perturbations pose greater challenges. However, models equipped with advanced reasoning capabilities tend to perform better on such tasks.
(4) \textbf{For queries requiring user profile understanding, LLMs show varying performance across different user interests and demographics.} 
For example, LLMs may be better at providing recommendations for popular interests and tend to perform more effectively for female users than for male users.
Our contributions are summarized as follows:
\begin{itemize}[leftmargin=*]
    \item \textbf{Novel Recommendation Paradigm:} 
    We introduce a new paradigm for next-generation recommender systems, exploring the effectiveness of LLMs acting as \emph{personalized recommendation assistants}  in an interactive and intelligent way. 
    As the first benchmark to integrate LLMs in this way, we transform the traditional recommendation paradigm into a more personalized and context-aware experience, setting a new standard for this field. 
    
    \item \textbf{Dataset Construction:} We present a comprehensive dataset called \ourname to simulate practical recommendation scenarios in the era of LLMs. 
    Our dataset encompasses approximately 30,000 complex user queries that require complex reasoning, including multiple conditions, multi-hop reasoning, misleading information, and diverse user profiles. 
    To the best of our knowledge, this is the first public dataset that can be used to evaluate the performance of LLM-based recommendation assistants in complex and reasoning-intensive recommendation settings.

    \item \textbf{Comprehensive Evaluation:} 
    We conduct extensive experiments with state-of-the-art LLMs and fine-tune open-source models for this task, analyzing their strengths and limitations in this task and providing actionable insights.

\end{itemize}

\section{The \ourname Benchmark}
In this section, we first present an overview of the \ourname benchmark. Then, we provide the construction methods of each data type included in our proposed dataset.

\subsection{Overview}

In practical applications, user demands can generally be categorized into two types, each requiring specialized capabilities from LLM-based recommendation assistants.
The first type is driven by specific conditions or constraints, such as asking for movies by a particular director or within a certain genre. 
This scenario demands strong reasoning capabilities from the recommendation assistant, as it requires the assistant to accurately interpret and apply the given constraints to generate appropriate recommendations. 
The second type does not involve specified conditions but requires recommendation assistants to generate recommendations based on an understanding of the user's profile (e.g., user interest, occupation, gender), which emphasizes the assistant's ability to interpret the profile and provide personalized suggestions. 
Given that these two scenarios demand different capabilities from the recommendation assistant, we construct two distinct query types to evaluate their performance: \textit{\firstname} and \textit{\secondname}. As shown in Table~\ref{tab:dataset_statistics}, our benchmark utilizes \textbf{Amazon-Book} and \textbf{Movielens-1M}\footnote{Unlike the 10M/20M versions, MovieLens-1M includes essential user demographic data.}. Below, we detail the construction of \firstname and \secondname queries based on these sources.

\subsubsection{\firstname Query}
The \textit{\firstname} Query simulates scenarios where users impose concrete constraints (e.g., specific directors or genres). The core challenge lies in accurately satisfying these conditions. To evaluate reasoning capabilities, we categorize this query into three types: \textit{Explicit Condition Query} (directly stated constraints), \textit{Implicit Condition Query} (vague inputs requiring multi-hop reasoning), and \textit{Misinformed Condition Query} (containing errors or typos).

\begin{itemize}[leftmargin=*]
    \item \textbf{\underline{Explicit Condition Query:}} 
    These queries explicitly specify the conditions that the recommended items should satisfy in the natural language format. 
    For example, the query, ``\emph{Can you recommend movies featuring Gwyneth Paltrow?}'', directly states the condition (``featured by Gwyneth Paltrow'') the recommendations must satisfy. Explicit condition queries are relatively straightforward because the constraints are clearly articulated.
    
    \item \textbf{\underline{Implicit Condition Query:}} Unlike Explicit Condition Query, \textit{Implicit Condition Query} involves incomplete specifications, demanding multi-hop reasoning to deduce target conditions. For example, requesting ``\emph{movies with the same cinematographer as Stay Hungry and Beethoven}'' requires inferring the unstated name (David Worth) first. This category serves as a more challenging benchmark for reasoning capabilities.

    \item \textbf{\underline{Misinformed Condition Query:}} Users may sometimes provide incorrect or misleading information about items in their queries. 
    For example, the query, ``\emph{I like `Avatar' directed by Spielberg. Please recommend similar movies directed by him.}'', incorrectly attributes the movie Avatar to Steven Spielberg. It is expected that recommendation assistants can first identify and correct this misinformation--recognizing that Avatar was not directed by Spielberg—before generating relevant recommendations. 
    This subcategory presents a significant challenge to the robustness of the recommendation assistant, requiring it to accurately identify inaccurate information in users' queries.
    \end{itemize}

\subsubsection{\secondname Query}
\label{sec:user_profile_based_query}
\begin{table}[t]
\vspace{-1.2 em}
  \centering
  \caption{Statistics of \ourname}
  \vspace{-1.1em}
  \resizebox{0.45\textwidth}{!}{ % Resizes the table to fit within the page width
    \begin{tabular}{ccccc} 
      \toprule
      \textbf{Major Category} & \multicolumn{1}{c}{\textbf{Sub Category}} & \textbf{Condition} & \textbf{Movie} & \textbf{Book} \\
      \midrule
      % \multirow{9}{*}{Condition-based Query}
      \multirow{9}[2]{*}{\makecell[c]{\firstname\\Query}}
      & \multirow{3}[2]{*}{\textit{\makecell[c]{Explicit\\Condition}}} & {1} & 2,225  & 2,260 \\
            &       & {2} & 2,346  & 2,604 \\
            &       & {3,4} & 426   & 271 \\
  \cmidrule(lr){2-5}    
   & \multirow{3}[2]{*}{\textit{\makecell[c]{Implicit\\Condition}}} & {1} & 1,753  & 1,626 \\
            &       & {2} & 1,552  & 2,071 \\
            &       & {3,4} & 344   & 213 \\
  \cmidrule(lr){2-5}    
  & \multirow{3}[2]{*}{\textit{\makecell[c]{Misinformed\\Condition}}} & {1} & 1,353  & 1,626 \\
            &       & {2} & 1,544  & 2,075 \\
            &       & {3,4} & 342   & 215 \\
  \midrule    \multicolumn{1}{c}{\multirow{2}[2]{*}{\makecell[c]{\secondname\\Query}}} & \textit{Interest-based} & - & 7,365  & 2,004 \\
            & \textit{Demographics-based} & - & 279   & 0 \\
  \midrule
  \multicolumn{3}{c}{\textbf{Queries in Total}} &\textbf{ 19,529} & \textbf{14,965} \\
  \midrule
  \multicolumn{3}{c}{\textbf{Number of Users}} &\textbf{ 6,036} & \textbf{48,535} \\
  \multicolumn{3}{c}{\textbf{Number of Items}} &\textbf{ 3,247} & \textbf{50,088} \\
  \multicolumn{3}{c}{\textbf{Number of User-Item Interactions}} &\textbf{ 29,459} & \textbf{55,424} \\
  
      \bottomrule
      \end{tabular}%
  }
  \label{tab:dataset_statistics}%
  \vspace{-1em}
\end{table}

In many real-world scenarios, users may not provide specific conditions to filter items but instead expect to be recommended personalized results based on their profile information, such as interests (e.g., interest in special effects) or demographic attributes (e.g., gender, occupation). 
This is because users often lack a clear understanding of their own needs and instead seek recommendations that intuitively align with their background or preferences. To address this common situation, we introduce \secondname Query, which simulates these scenarios by requiring the recommendation assistant to generate personalized suggestions based on user profile information. We further divide \secondname Query into \textit{Interest-Based} and \textit{Demographics-Based} Query, since interests reflect users' dynamic preferences, while demographics capture more stable attributes like gender or occupation that influence their needs.
\begin{figure*}[t]
\vspace{-1em}
    \centering
    \includegraphics[width=0.75\linewidth]{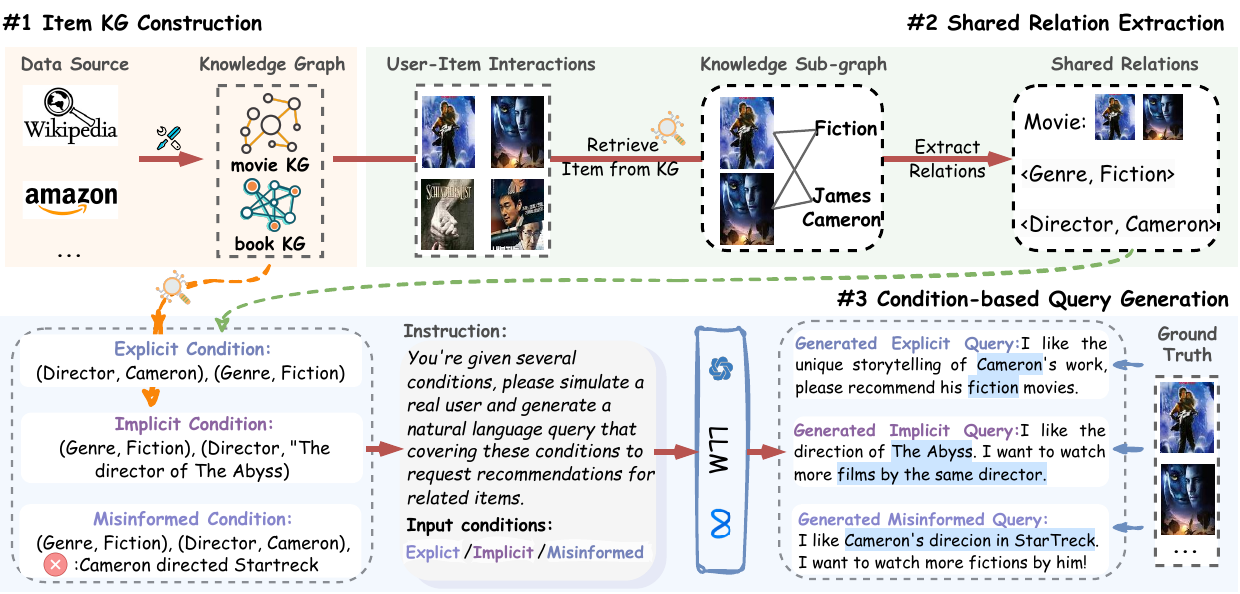}
    \vspace{-1em}
    \caption{Overview of the pipeline of constructing \textit{\firstname Query}. }
    \label{fig:conditionbased-construction}
\vspace{-1.5em}
\end{figure*}

\begin{itemize}[leftmargin=*]
    \item 
    \textbf{\underline{{Interest-based Query}:}} These queries reflect user interests that can't be clearly defined by specific conditions. For example, users who watched ``\emph{The Tree of Life}'' and ``\emph{2046}'' in history may prefer movies with unique storytelling and visual beauty. 
    To capture such interests, we create queries like, ``\emph{I just watched `The Tree of Life' and loved its {poetic storytelling and visual design}. Can you recommend other movies with a similar tone?}'' Such queries don't have explicit conditions but require the personalized recommendation assistant to understand and suggest items based on the user's personal interest. 
    
    \item \textbf{\underline{{Demographics-based Query}}}: These queries focus on how users' demographic attributes influence recommendations. 
    Different users, based on their age, gender, or occupation, may have distinct preferences and needs.
    For example, philosophy professors might be interested in films that explore moral dilemmas.
    By leveraging these patterns, we can craft queries such as: ``\emph{I'm a philosophy professor looking for movies with complex themes and moral dilemmas. Can you recommend any?}'' 
    By aligning with real-world demographic behaviors, these queries can mirror the personalized recommendations that a personalized recommendation assistant should provide, ensuring that the suggestions are relevant and tailored to the user's background and identity.

\end{itemize}

\subsection{\firstname Query Construction}
\label{sec:condition query construction}

User queries often express preferences through conditions corresponding to item attributes, such as filtering movies by genre, director, or actor. However, due to the intricate relationships between items and their attributes, it is challenging to generate realistic, diverse queries that reflect how users combine these attributes. 
It is worth noting that Knowledge Graph (KG), as a structured representation, explicitly connects items (e.g., movies) to their attributes (e.g., directors) and captures interdependencies between them (e.g., ``director X frequently collaborates with actor Y''). 
Therefore, we propose leveraging KG to extract meaningful conditions and generate queries aligned with real-world user behavior.
Specifically, our approach contains three key steps: \textit{Item Knowledge Graph (Item KG) construction}, \textit{shared relation extraction}, and \textit{query generation}, as shown in Figure~\ref{fig:conditionbased-construction}. We first build the Item KG, which links items to their attributes.
Next, we identify item sets from the user's interaction history that share the same attributes (e.g., movies directed by Christopher Nolan), forming the conditions for the query. 
Finally, we adopt LLMs to simulate users and generate queries based on these extracted conditions. The item sets that satisfy the conditions can serve as the ground truth for evaluation.

\subsubsection{Item KG Construction}
The Item KG forms the basis for generating realistic \firstname Query in \ourname. We take advantage of different KGs for the movie and book domains:
\begin{itemize}[leftmargin=*]
    \item Movies: We extract data from Wikipedia, focusing on 7 key attributes like directors, actors, composers, and genres. Each movie node is linked to these attributes.
    \item Books: We use metadata from the Amazon Book Dataset, connecting attributes like authors and categories to book nodes.
\end{itemize}
Once the KGs are built, we combine them with traditional recommendation datasets, Movielens-1M and Amazon-Book, to generate conditions and queries step by step.

\subsubsection{Shared Relation Extraction}

Shared relations are common attributes found among items in a user's interaction history. For example, if a user has watched multiple movies by the same director or starring the same actor, these shared attributes help generate \firstname Queries.
Given a user \( u \) with an interaction history \( \mathcal{H}_u = \{i_1, i_2, \ldots, i_k\} \), we employ a KG retrieval function \( \mathcal{R} \) to identify shared attributes (relations) across subsets of items in \( \mathcal{H}_u \). Each shared relation is defined as a tuple \( (r, t) \), where \( r \) is the type of relation (e.g., ``directed by'', ``starring'') and \( t \) is the target value (e.g., name of director). The extraction process produces multiple groups of shared relations and their corresponding subsets of items, which can be represented as follows:
\abovedisplayskip=0pt
        \begin{align}
        \mathcal{R}(\mathcal{H}_u, KG) = 
        \left\{ \left(\mathcal{G}_{\text{sub}}, \mathcal{C}_{\text{shared}} \right) \; \middle| \;
        \mathcal{G}_{\text{sub}} \subseteq \mathcal{H}_u, \right. \notag \\
        \left. \mathcal{C}_{\text{shared}} = \{(r_1, t_1), (r_2, t_2), \ldots \} \right\}.
        \end{align}
\belowdisplayskip=0pt
\vskip -0.6em
    Here, \( \mathcal{G}_{\text{sub}} \) represents a subset of items sharing the same relations \( \mathcal{C}_{\text{shared}} \).  The extracted shared relations will later serve as conditions for query generation.

\begin{figure*}[ht!]
\vspace{-1em}
    \centering
    \includegraphics[width=0.90\linewidth]{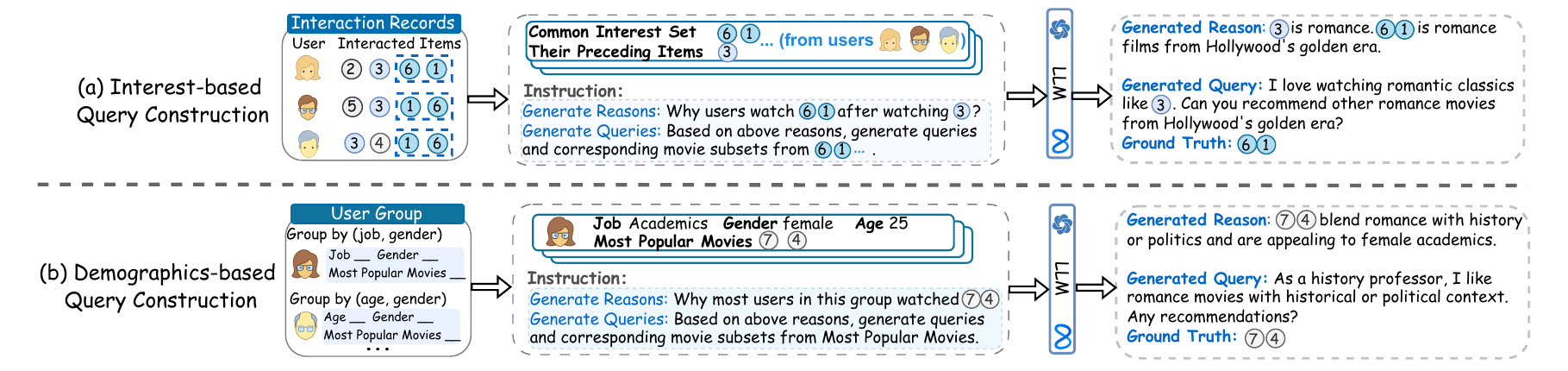}
    \vspace{-1.5em}
    \caption{Overview of the pipeline of constructing \textit{\secondname Query}.}
    \label{fig:collaborativebased_construction}
    \vspace{-1em}
\end{figure*}

\subsubsection{Query Generation}
After extracting shared relations from the user's interaction history, we generate three types of \firstname Query: explicit, implicit, and misinformed. The steps to generate these conditions and final queries are as follows:

\begin{itemize}[leftmargin=*]
    \item \textbf{\underline{Explicit Condition Construction:}} For explicit conditions, we directly adopt the shared relations extracted from the user's interaction history. Specifically, we set the explicit condition \(\mathcal{C}_{\text{explicit}} \ = \mathcal{C}_{\text{shared}} \). This means the conditions are directly derived from the shared attributes (e.g., director, genre) associated with the items the user has interacted with.
    
    \item \textbf{\underline{Implicit Condition Construction:} }For implicit conditions, instead of directly stating the specific value (e.g., the director's name), we leverage the KG to describe them indirectly through related items. For example, instead of directly referencing James Cameron, we might use \textit{``the director of `The Abyss' ''}.  
    Specifically, for a shared relation \((r_m, t_m) \in \mathcal{C}_{\text{shared}}\) (e.g., (director, Cameron)), we replace the value that needs to be inferred, \(t_m\) (e.g., Cameron), with an indirect reference that describes \(t_m\)'s relation with another item \(i_k\) (e.g., The Abyss) from the KG. Mathematically, this can be expressed as:
\abovedisplayskip=0pt
\begin{align}
i_k \in \{i \in \mathcal{I} \mid (t_m) \xrightarrow{r_m} (i) \text{ in KG}\},
\end{align}
\belowdisplayskip=0pt

where \(\mathcal{I}\) is the set of all items in the KG. The resulting implicit condition is then:
\abovedisplayskip=0pt
\begin{align}
\mathcal{C}_{\text{implicit}} = (r_i, \text{ref}(i_k)),
\end{align}
\belowdisplayskip=0pt

where \(\text{ref}(i_k)\) denotes a textual reference to \(i_k\). For example, a shared relation like (director, Cameron) can be transformed into an implicit condition (director, ``the director of `The Abyss' ''), where ``The Abyss'' (\(i_k\)) is retrieved from the KG as being connected to ``Cameron'' (\(t_0\)) via the ``director'' relation (\(r_0\)).
    \item \textbf{\underline{Misinformed Condition Construction:}} For misinformed conditions, we intentionally introduce factual error into the conditions, simulating real-world user errors where they might mistakenly reference attributes or relationships. To generate a misinformed condition, for a shared relation \((r_i, t_i) \in \mathcal{C}_{\text{shared}}\), we randomly select one or more items \(\{i_1, i_2, \ldots, i_m\} \in \mathcal{I}\) from the KG that do not have the specified relationship $r_i$ with $t_i$ (i.e., \(i_k \not\longrightarrow t_i\) for each \(i_k \in \{i_1, i_2, \ldots, i_m\}\)). We then construct the condition with ``error info'' that falsely claims these items are related to \(t_i\) through \(r_i\):
    \begin{align}
        \mathcal{C}_{\text{misinformed}} = (r_i, t_i, \text{error info: } \{i_1, i_2, \ldots, i_m\} \xrightarrow{r_i} t_i),
\end{align}
   
    where the set \(\{i_1, i_2, \ldots, i_m\}\) contains items from the KG that do not actually share the relationship \(r_i\) with \(t_i\).
    For example, a misinformed condition can be (director, Cameron, error info: \textit{Cameron is the director of `Startreck'}).
    \item \textbf{\underline{Query Generation:}} For each condition type, we apply GPT-4o~\cite{hurst2024gpt} to generate corresponding  queries:
\begin{align}
                q = LLM(\mathcal{C}, \mathcal{P}),
\end{align}
     where $\mathcal{P}$ is the prompt.
    For each generated query \( q \), we construct a sample consisting of:
    \begin{align}
    (q, \mathcal{H}_u', \mathcal{G}_q),
    \ \text{where:} \
    \mathcal{H}_u' = \mathcal{H}_u \setminus \mathcal{G}_q, \\ \quad \mathcal{G}_q = \{i \in \text{KG} \mid \mathcal{C}_{\text{shared}} \subseteq \mathcal{A}_i\}.
    \end{align}
    
    Here, \( \mathcal{H}_u' \) is the user's interaction history excluding items in \( \mathcal{G}_q \), and \( \mathcal{G}_q \) contains the ground truth items satisfying the query conditions. \( \mathcal{A}_i \) represents the attributes of item \( i \) in the KG.

\end{itemize}

% \vspace{-1.5em}
\subsection{\secondname Query Construction}
\label{sec:collaborative query construction}

In addition to providing specific constraints, users may provide only basic personal information such as watching history or demographic attributes (e.g., age, occupation). To evaluate the LLM recommendation assistant’s ability to handle such scenarios, we introduce \secondname Query, which aims to generate personalized recommendations based on user profile information without requiring specific filtering conditions. As detailed in Section~\ref{sec:user_profile_based_query}, \secondname Query contains two subcategories: \textit{Interest-based Query} and \textit{Demographics-based Query}. The construction process is illustrated in Figure~\ref{fig:collaborativebased_construction} .

\subsubsection{Interest-based Query}
These queries are designed to capture user interests that are not explicitly defined by specific conditions but are inferred from user interactions. To effectively simulate real-world scenarios, we focus on capturing the collective interests that emerge from shared behaviors from multiple users, which are more representative of general preferences and trends. For example, if many users who watched ``The Dark Knight'' also watched ``Inception'', this suggests a clear interest in action and suspense. In this way, we can craft queries that reflect collective user interests derived from shared behaviors.
 Specifically, we first identify \textbf{Common Interest Set}, which are sets of items that frequently interact consecutively by multiple users. Let $\mathcal{H}_u$ denote the interaction history of user $u$, and $\mathcal{H} = \{\mathcal{H}_u \mid u \in \mathcal{U}\}$ denote the set of all users' interaction histories. The common interest set can be defined as:
 % \vspace{-0.7em}
\begin{align}
    S = \{s \mid \exists u \in \mathcal{U}, s \subseteq \mathcal{H}_u, f(s) \geq \theta\},
\end{align}
where $f(s)$ is the frequency of sequence $s$ across all users' histories, and $\theta$ is a predefined frequency threshold.

Once common interest sets are identified, we proceed to extract the \textbf{preceding item sequences}—that is, the sequences of items that commonly appear before the items of the common interest set. Let $P(s)$ denote the set of preceding item sequences for target sequence $s$. The preceding sequences can be defined as:
\begin{align}
P(s) = \{p \mid \exists u \in \mathcal{U}, p \prec s \subseteq \mathcal{H}_u\},
\end{align}
where $p \prec s$ denotes that sequence $p$ appears immediately before sequence $s$ in the interaction history $\mathcal{H}_u$.

% Table generated by Excel2LaTeX from sheet 'overall performance'
\begin{table*}[h!]
\vspace{-1em}
  \centering
  \caption{Performance on \firstname Query. 
  The \textcolor{backred!50}{\rule{0.5cm}{0.2cm}}/\textcolor{backblue!75}{\rule{0.5cm}{0.2cm}} means the best/the second-best result. }
  \vspace{-1em}
  % \begin{small}
  \resizebox{0.95\textwidth}{!}{
    \begin{tabular}{cccccc|cccc|cccc|ccc}
    \toprule
    \multirow{2}[2]{*}{Domain} & \multirow{2}[2]{*}{Model} & \multicolumn{4}{c|}{Explicit Condition (Easy)} & \multicolumn{4}{c|}{Implicit Condition (Medium)} & \multicolumn{4}{c}{Misinformed Condition (Hard)} & \multicolumn{3}{c}{Average} \\
          &       & P$\uparrow$ & R$\uparrow$ & CMR$\uparrow$   & FTR$\downarrow$   & P$\uparrow$ & R$\uparrow$ & CMR$\uparrow$   & FTR$\downarrow$   & P$\uparrow$ & R$\uparrow$ & CMR$\uparrow$   & FTR$\uparrow$ & P$\uparrow$ & R$\uparrow$ & CMR$\uparrow$  \\
    \midrule
    \multirow{7}[2]{*}{Movie} & GPT-4o-mini & 0.185 & 0.322 & 0.531 & 0.009 & 0.083 & 0.167 & 0.198 & 0.017 & 0.028 & 0.060  & \cellcolor{backred!50}{0.153} & 0.104 & 0.099 & 0.183 & 0.294 \\
  
          & GPT-4o & \cellcolor{backred!50}{0.308} & 0.408 & \cellcolor{backred!50}{0.714} & 0.016 & \cellcolor{backblue!75}{0.145} & 0.224 & \cellcolor{backblue!75}{0.301} & 0.021 & 0.019 & 0.039 & 0.106 & \cellcolor{backred!50}{0.270}  & \cellcolor{backred!50}{0.157}  & 0.224 & \cellcolor{backblue!75}{0.374}  \\
          
          & Gemini & \cellcolor{backblue!75}{0.256} & 0.408 & 0.644 & 0.052 & 0.104 & 0.206 & 0.203 & 0.014 & 0.024 & 0.049 & 0.076 & 0.030  & 0.128 & 0.221 & 0.308  \\
          \
          & Claude &   0.201    & \cellcolor{backblue!75}{0.422} & 0.658 & 0.014 & 0.105 & 0.269 & 0.281 & {0.011} & \cellcolor{backblue!75}{0.033} & \cellcolor{backred!50}{0.079} & \cellcolor{backblue!75}{0.128} & 0.087 & 0.069 & 0.183 & 0.277 \\
          
          & DeepSeek-V3 & 0.190 & 0.401 & 0.621 & \cellcolor{backred!50}{0.001}  & 0.090 & {0.260} & 0.217 & \cellcolor{backred!50}{0.001} & 0.027 & \cellcolor{backblue!50}{0.078} & 0.105 & 0.013 & 0.102 & \cellcolor{backblue!75}{0.246} & 0.314 \\
          
          & DeepSeek-R1  & 0.224 & \cellcolor{backred!50}{0.447} & \cellcolor{backblue!75}{0.651} & \cellcolor{backred!50}{0.001} & \cellcolor{backred!50}{0.197} & \cellcolor{backred!50}{0.463}  & \cellcolor{backred!50}{0.496} & \cellcolor{backblue!75}{0.005} & 0.024 & {0.068} & 0.096 & 0.024 & \cellcolor{backblue!75}{0.148} & \cellcolor{backred!50}{0.326} & \cellcolor{backred!50}{0.414} \\
          
          & Llama-3.1-70B & 0.238 & 0.342 & 0.609 & \cellcolor{backblue!75}{0.003} & 0.097 & 0.164 & 0.210  & 0.012 & \cellcolor{backred!50}{0.037} & 0.050  & 0.116 & \cellcolor{backblue!75}{0.109} & 0.124 & 0.185 & 0.312 \\
\cmidrule{1-17} 

    \multirow{7}[2]{*}{Book} & GPT-4o-mini & 0.059 & 0.159 & 0.475 & \cellcolor{backred!50}{0.003} & 0.035 & 0.081 & 0.446 & \cellcolor{backred!50}{0.003} & 0.013 & {0.038} & \cellcolor{backred!50}{0.581} & 0.044 & 0.036 & 0.093 & 0.501\\
          & GPT-4o & \cellcolor{backred!50}{0.088} & 0.192 & 0.567 & 0.027 & \cellcolor{backblue!75}{0.057} & 0.133 & \cellcolor{backblue!75}{0.472} & 0.021 & 0.011 & 0.024 & \cellcolor{backblue!75}{0.500}  & \cellcolor{backred!50}{0.445} & \cellcolor{backred!50}{0.052} & 0.116 & \cellcolor{backblue!75}{0.513}\\
          & Gemini & \cellcolor{backblue!75}{0.076} & \cellcolor{backblue!75}{0.221} & \cellcolor{backblue!75}{0.623} & 0.011 & 0.035 & 0.135 & 0.319 & \cellcolor{backblue!75}{0.013} & {0.014} & 0.044 & 0.274 & 0.072 & 0.042 & 0.133 & 0.405\\
          
          & Claude &  0.054     &  {0.193}     &   0.608    &    0.010   &    0.043   &    {0.161}   &   \cellcolor{backred!50}{0.515}    &   0.010    & \cellcolor{backred!50}{ 0.020}     &    \cellcolor{backred!50}{0.068}   &    0.444   & 0.056  & 0.039 & \cellcolor{backblue!75}{0.141} & \cellcolor{backred!50}{0.522}\\
          
          & DeepSeek-V3 & 0.040 & 0.124 & \cellcolor{backred!50}{0.667} & \cellcolor{backblue!75}{0.008} & 0.056 & \cellcolor{backblue!75}{0.190} & 0.385 & 0.005 & {0.014 }& {0.047} & 0.230 & 0.066 & 0.037 & 0.120 & 0.351\\
          
          & DeepSeek-R1  & 0.072 & \cellcolor{backred!50}{0.230} & 0.471 & 0.018 & \cellcolor{backred!50}{0.060} & \cellcolor{backred!50}{0.194} & 0.167 & 0.031 & \cellcolor{backred!50}{0.015} & \cellcolor{backblue!75}{0.051} & 0.333 & 0.097 & \cellcolor{backblue!75}{0.049} & \cellcolor{backred!50}{0.159} & 0.324\\
          
          & Llama-3.1-70B & 0.073 & 0.170  & 0.542 & 0.022 & 0.038 & 0.082 & 0.470  & 0.052 & \cellcolor{backblue!75}{0.014} & 0.028 & 0.178 & \cellcolor{backblue!75}{0.158} & 0.042 & 0.093 & 0.397\\

    \bottomrule
    \end{tabular}%
  \label{tab:condition_performance}%
% \end{small}
}
\vspace{-1em}
\end{table*}%

Based on the identified common interest set and preceding item sequences, we employ an LLM to analyze the patterns and infer the reasoning behind multiple users' common interests. The generated query aims to reflect both the contextual features of the items and the common interests of users.

% \vspace{-1em}
\subsubsection{Demographics-based Query}
Individuals with similar demographics often share preferences; for instance, children typically favor animated films (e.g., Disney), whereas teenagers may prefer action or superhero movies. Leveraging this, we group users by attribute permutations (e.g., students aged 18–24) and identify their top-consumed items. We then input these demographics and item lists into an LLM to synthesize queries that capture group-specific patterns. The identified popular items serve as the ground truth for evaluation.

\subsection{Data Filtering and Selection}
For Condition-based Query, we use string matching to check if the provided information (e.g., actor, director) appears in the query and filter out queries that omit this information. For User Profile-based Query, we manually verify the queries and filter out those that are too broad or unreasonable.

\section{Experiments}

\subsection{Experiments Settings}

\subsubsection{Baselines}
We evaluate several widely used models, including GPT-4o (2024-08-06)~\cite{hurst2024gpt}, GPT-4o-mini (2024-07-18)~\cite{hurst2024gpt}, Llama (Llama-3.1-70B-Instruct)~\cite{dubey2024llama}, Gemini (gemini-1.5-pro-002)~\cite{team2023gemini}, 
Claude (claude-3-5-sonnet-20241022), DeepSeek-V3~\cite{liu2024deepseek}, and DeepSeek-R1~\cite{guo2025deepseek} on our proposed dataset. However, we were unable to test the GPT-o1~\cite{jaech2024openai} model because it flagged our prompts as violating its usage policy. In addition, we further perform full-parameter supervised fine-tuning (SFT) and reinforcement fine-tuning (RFT, based on the GRPO~\cite{shao2024deepseekmath} algorithm) on the Qwen-2.5-3B-Instruct and LLaMA-3.2-3B-Instruct models to explore their potential performance gains under task-specific optimization.

\subsubsection{Evaluation Metrics}
We utilize the below metrics to evaluate LLM performance. To simulate real-world scenarios where users typically do not predefine the number of recommendations they need, we do not specify a fixed number of recommendations K in the testing prompts; however, we also include several experiments with a fixed K in Section~\ref{sec:exp_condition} for reference.

\begin{itemize}[leftmargin=*,itemsep=0pt, parsep=0pt, topsep=0pt, partopsep=0pt]
    \item \textbf{Precision:} The proportion of recommended items that are actually relevant to the user's query. 
    \item \textbf{Recall~\cite{he2020lightgcn}:} The proportion of relevant items that are successfully recommended out of all relevant items available.
    \item \textbf{Condition Match Rate (CMR):} The percentage of recommended items that meet the specified conditions. For \firstname Queries, it is important to check if the model’s recommendations strictly align with the given conditions, as items that don’t meet these are likely unsatisfactory. Traditional metrics like Precision and Recall are insufficient for evaluating this alignment. Therefore, we propose CMR to achieve this goal.
    \item \textbf{Fail to Recommend (FTR):} This metric measures the proportion of queries where the model fails to generate recommended items. A lower FTR is preferred, indicating the model can generate a list of items. However, in Misinformed \firstname Query, a higher FTR shows the model's ability to identify incorrect information, thereby avoiding wrong recommendations.
\end{itemize}

\subsection{Results on \firstname Query}
\label{sec:exp_condition}

\textbf{\underline{Observation 1.} For different LLMs, GPT-4o and DeepSeek-R1 outperform other models.} As shown in Table~\ref{tab:condition_performance}, GPT-4o achieves the highest Precision and the second-highest CMR average score on both the movie and book datasets, while DeepSeek-R1 ranks first in Recall and second in Precision. Notably, models with advanced reasoning capabilities, such as DeepSeek-R1, excel in handling query types that require reasoning. For example, in Implicit Condition Query, DeepSeek-R1 exhibits a significantly smaller drop in performance compared to other models when transitioning from explicit to implicit conditions. This positions it as the leader in Implicit Condition Query performance. It suggests that the advanced reasoning ability enables the LLM-based assistant to infer implicit attributes not explicitly mentioned in the conditions, leading to more accurate recommendations. There is a case study in \textbf{Appendix~\ref{app:case study}}.

\noindent \textbf{\underline{Observation 2.} For different query types, model performance decreases with increasing query difficulty.}
In both the movie and book datasets, most LLMs perform best on Explicit Condition Query, outperforming the other two query types. This suggests that LLMs are better at handling queries that clearly state conditions. In contrast, 
implicit conditions present a greater challenge because models must infer constraints about items that are not explicitly stated. It is worth noting that models with strong reasoning capabilities like Deepseek-R1 maintain strong performance, suggesting benefits from its reasoning capability. In addition, the misinformed condition proves most difficult, with the lowest recall and CMR, indicating frequent failures in providing personalized recommendations when queries contain misleading information. Moreover, higher FTR rates under this condition suggest that LLM-based assistants may leverage general knowledge to detect misinformation in queries and avoid directly generating inaccurate recommendations.

\begin{figure}[t]
\vspace{-1.2em}
    \centering    \includegraphics[width=0.9\linewidth]{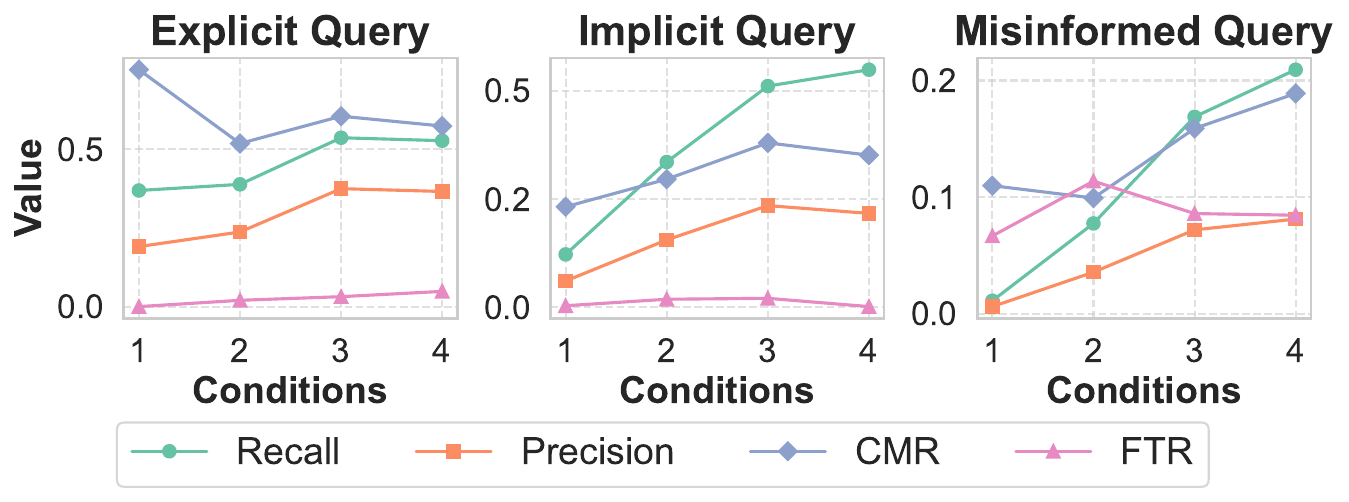}
    \vspace{-1.5em}
    \caption{The effect of number of conditions.}
    \label{fig:condition_linechart}
    \vspace{-1.5em}
\end{figure}

\noindent \textbf{\underline{Observation 3.} 
For different numbers of conditions in queries, Precision and Recall improve with more conditions, while CMR declines for Explicit Condition Query and rises for others. } As Figure~\ref{fig:condition_linechart} illustrated, we take the movie dataset as an example and investigate the impact of varying the number of conditions in queries on model performance. 
We note that additional conditions narrow the recommendation scope, thus improving the recommendation accuracy.
 For example, the set of candidates for \textit{``Action movies by Nolan released after 2010''} is much smaller than \textit{``movies by Nolan''}, reducing ambiguity and improving the relevance of the recommendations. 
The impact of condition quantity on CMR metrics varies across query types. In Explicit Condition Query, CMR is highest with a single condition, as adding more conditions increases the difficulty of satisfying all constraints, leading to reduced coverage. In contrast, for Implicit and Misinformed Condition Query, additional conditions provide more context, allowing the model to better infer implicit conditions and reduce the impact of factual error in the conditions. As a result, the CMR metric improves with more conditions in these query types.

\noindent  \textbf{\underline{Observation 4.} 
\begin{figure}[b]
    \vspace{-1.2em}
    \centering
    \includegraphics[width=\linewidth]{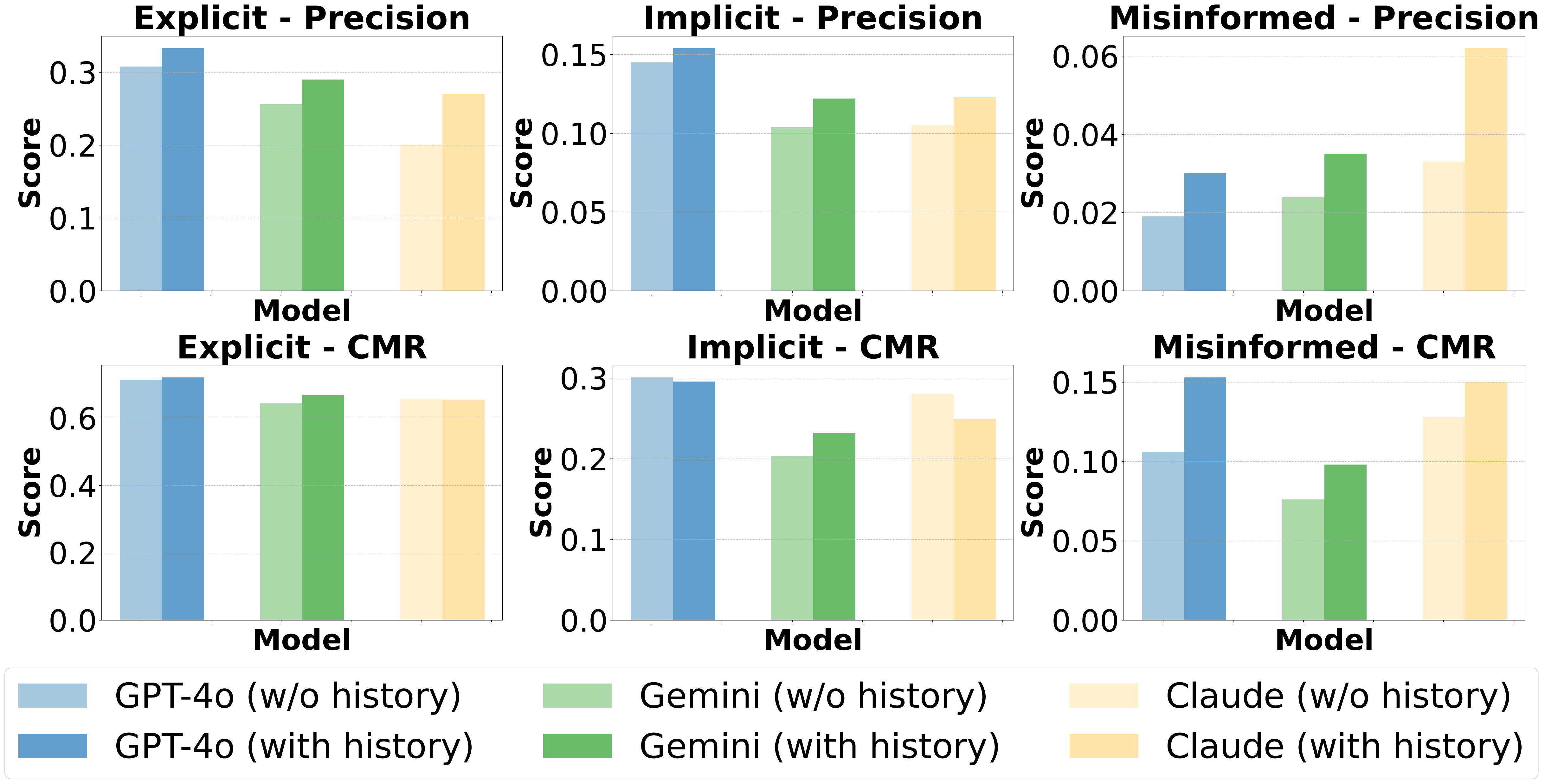}
    \vspace{-2em}
    \caption{The effect of incorporating user-item history.}
    \label{fig:history_effect}
    % \vspace{-2.5em}
\end{figure}
For the effect of user-item interaction history, incorporating them enhances recommendation quality in two key ways.}
The effect of incorporating user-item interaction history is shown in Figure~\ref{fig:history_effect}.
First, incorporating user-item history improves Precision across all query types by generating a more personalized recommendation list. The reason can be that queries with limited conditions (e.g., “movies directed by Nolan”) often yield a large candidate pool. By leveraging historical interactions, the model can prioritize select items that align with the user’s preferences and satisfy the query conditions, effectively filtering irrelevant candidates. 
% We provide examples of both using and not using history in the \textbf{Appendix~\ref{app:case_study_history}}.

% For the effect of user-item interaction history, incorporating user-item history improves Precision across all query types but only enhances the CMR metric for Misinformed condition-based queries. } The effect of incorporating user-item interaction history is shown in Figure~\ref{fig:history_effect}.
% Incorporating user-item interaction history consistently improves Precision by generating a more focused recommendation list. 
% \jn{This is intuitive, as a limited number of conditions typically results in a larger pool of items that meet these conditions. By incorporating the users' historical data, the model can better understand their preferences. Consequently, it can more effectively select items that are not only condition-satisfied but also closely aligned with user preferences.} \sj{***The descriptions are too spoken; we need to write more academically. ***} We provide examples of both using and not using history in the appendix~\ref{app:case_study_history}.

Second, incorporating user history does not always improve CMR. By analyzing the output of the LLM assistant, we note that adding history can also introduce “distractor” items that reduce the model's strict condition adherence. For example, the LLM assistant may recommend ``The Prestige'' for ``Nolan films'' due to the user’s fantasy genre bias in history. This suggests that incorporating history may divert the model's attention, making it more challenging to strictly adhere to the given conditions. These results highlight an important trade-off: while incorporating history can enhance the model's understanding of user preferences, it may also hinder the model's ability to closely align with the specific requirements in the query. Consequently, balancing the influence of historical data with the need for precise condition adherence is crucial for optimizing performance in recommendation assistants.

\begin{table}[t]
  \centering
  \vspace{-1em}
  \caption{Results on Condition-based Query with K=5}
   \vspace{-1em}
   
  \begin{small}
  \resizebox{0.48\textwidth}{!}{
    \begin{tabular}{ccc|cc|cc|cc}
    \toprule
    \multirow{2}[2]{*}{Model} & \multicolumn{2}{c|}{Explicit} & \multicolumn{2}{c|}{Implicit} & \multicolumn{2}{c|}{Misinformed} & \multicolumn{2}{c}{Average} \\
           & Recall & CMR    & Recall & CMR   & Recall & CMR  & \multicolumn{1}{c}{Recall} & \multicolumn{1}{c}{CMR} \\
    \midrule

    GPT-4o & 0.436 & 0.657  & 0.245 & 0.266   & 0.052 & 0.077  & 0.244 & 0.333 \\
    Gemini & 0.416 & 0.607 & 0.205 & 0.195  & 0.048 & 0.062  & 0.223 & 0.288 \\
    Claude & 0.436 & 0.647  & 0.247 & 0.277   & 0.075 & 0.105  & 0.253 & 0.343 \\
    \bottomrule
    \end{tabular}%
        }
\end{small}
\vspace{-1.5em}
  \label{tab:k_experiment}%
\end{table}%

\noindent  \textbf{\underline{Observation 5.} Requiring the LLM to generate K recommendations aligns with previous observations. }
For Condition-based Query, there are samples where the number of items that satisfy the conditions in the query may be fewer than K. In such cases, requesting the model to output K recommendations could lead to suboptimal results. Therefore, in the main experiment presented in Table~\ref{tab:condition_performance}, we did not require the model to generate K recommendations. However, to further explore this aspect, we also conducted additional experiments in which the LLM was instructed to output K items. The results, shown in Table~\ref{tab:k_experiment}, are based on the movie dataset and align with previous findings: LLMs perform best on Explicit Condition Queries, with performance gradually declining for Implicit Condition and Misinformed Condition Queries.

\begin{table}[b]
  \centering
  \vspace{-1.5em}
  \caption{Fine-tuning performance on Condition-based Query}
  \vspace{-1.3em}
  \small
   \resizebox{0.45\textwidth}{!}{
    \begin{tabular}{cccccc}
    \toprule
    \multicolumn{1}{c}{Domain} & \multicolumn{1}{c}{Model} & Fine-tuning & \multicolumn{1}{c}{Explicit} & \multicolumn{1}{c}{Implicit} & \multicolumn{1}{c}{Misinformed} \\
    \midrule
    \multirow{6}[4]{*}{Movie} & \multirow{4}[2]{*}{Qwen-2.5-3B} & -     &  0.019     & 0.002      &  0.004 \\
          &       & SFT   & \underline{0.391}      & \underline{0.255}      & \underline{0.296} \\
          &       & RFT    &  0.178     & 0.119      & 0.089  \\
           &       & SFT+RFT    & \textbf{ 0.410 }    & \textbf{0.262}      & \textbf{0.307}  \\
\cmidrule{2-6}          & \multirow{4}[2]{*}{Llama-3.2-3B} & -     &    0.018   & 0.003      &  0.003\\
              &       & SFT   &   \underline{0.285}    &  \underline{0.176}     & \underline{0.186} \\
          &       & RFT    &  0.191     & 0.124      & 0.080 \\ 
          &       & SFT+RFT    &  \textbf{0.293}     & \textbf{0.181}      & \textbf{0.195}  \\
        
    \midrule
    \multirow{6}[4]{*}{Book} & \multirow{4}[2]{*}{Qwen-2.5-3B} & -     &   0.012    & 0.007      & 0.002 \\
          &       & SFT   &   \underline{0.136}    &  \underline{0.251}      & \underline{0.249} \\
          &       & RFT    &  0.055     & 0.090      & 0.081 
          \\
          &       & SFT+RFT    &  \textbf{0.139}    & \textbf{0.262}      & \textbf{0.264}  \\
\cmidrule{2-6}          & \multirow{4}[2]{*}{Llama-3.2-3B} & -     &  0.006     & 0.004      & 0.002 \\
          &       & SFT   &  \underline{ 0.088}    &   \underline{0.221}    &  \underline{0.181} \\
          &       & RFT    &   0.056    & 0.092      & 0.090 \\
          &       & SFT+RFT    & \textbf{ 0.112}     & \textbf{0.237}     & \textbf{0.190}  \\
    \bottomrule
    \end{tabular}%
  }
  \vspace{-1.5em}
  \label{tab:condition-finetune}%
\end{table}%

\noindent  \textbf{\underline{Observation 6.} For the performance of fine-tuning, both SFT and RFT significantly enhance the model’s recommendation performance.} Given the recent interest in using RFT to improve model reasoning capabilities~\cite{shao2024deepseekmath,xu2025towards}, we aimed to explore its potential in this context. We applied full-parameter SFT and RFT (based on the GRPO~\cite{shao2024deepseekmath} algorithm) to fine-tune the Qwen-2.5-3B-Instruct and LLaMA-3.2-3B-Instruct models, and evaluated recall performance across different query types. As shown in Table~\ref{tab:condition-finetune}, both fine-tuning paradigms notably improved the models' recommendation performances. However, models trained solely with RFT performed worse than those trained with SFT. This performance gap may be attributed to the lack of a warm-up phase for reasoning, which is essential for effective exploration during reinforcement learning~\cite{cai2025much}. To validate this, we further divided the training process into two stages: first applying SFT as a warm-up, followed by RFT training. Experiment results demonstrate that this two-stage approach further improved LLMs' performance. For example, in the movie's explicit condition-based queries, the SFT+RFT model achieved a recall of 0.410, outperforming the SFT-only model (0.391) and the RFT-only model (0.178).

\subsection{Results on \secondname Query}
% \vspace{-1em}

\noindent \textbf{\underline{Observation 7.} For different LLMs, Gemini-1.5 Pro and DeepSeek-R1 demonstrated better performance compared to other models.}
As shown in Table~\ref{tab:collaborative_performance}, larger and more advanced models, such as DeepSeek-R1 and Gemini-1.5-Pro, consistently achieve higher Precision and Recall. In contrast, smaller models like GPT-4o-mini tend to lag behind. From the FTR (Failure to Recommend) metrics perspective, most models exhibit a low failure rate, indicating their ability to effectively follow instructions and generate recommendation lists in the required format.
\begin{table}[t]
\vspace{-0.5em}
  \centering
  \vspace{-0.5em}
  \caption{Performance on \secondname Query.   The \textcolor{backred!50}{\rule{0.5cm}{0.2cm}}/\textcolor{backblue!75}{\rule{0.5cm}{0.2cm}} means the best/the second-best result, respectively.}
  \vspace{-1em}
  \resizebox{0.48\textwidth}{!}{
    \begin{tabular}{ccccc|ccc|ccc}
    \toprule
    \multirow{2}[2]{*}{} & \multirow{2}[2]{*}{Model} & \multicolumn{3}{c|}{Interest-based Query} & \multicolumn{3}{c|}{Demographics-based Query} & \multicolumn{3}{c}{Average} \\
          &       & P$\uparrow$ & R$\uparrow$ & FTR$\downarrow$   & P$\uparrow$ & R$\uparrow$ & FTR$\downarrow$ & P$\uparrow$ & R$\uparrow$ & FTR$\downarrow$ \\
    \midrule
    \multirow{7}[2]{*}{\rotatebox{90}{Movie}} 
    & GPT-4o-mini & 0.013 & 0.058 & 0.000 & 0.018 & 0.054 & 0.000 & 0.015 & 0.056 & {0.000}\\
    
          & GPT-4o & \cellcolor{backblue!75}{0.018} & 0.067 & 0.001 & \cellcolor{backred!50}{0.021} & {0.059} & 0.000 & \cellcolor{backred!50}{0.020} & 0.063 & {0.000}\\
          
          & Gemini & \cellcolor{backred!50}{0.019} & {0.072} & 0.007 & \cellcolor{backblue!75}{0.019} & \cellcolor{backblue!50}{0.063} & 0.000 & \cellcolor{backblue!75}{0.019} & \cellcolor{backblue!75}{0.072} & 0.004\\
          
          & Claude & 0.015 & \cellcolor{backred!50}{0.082} & 0.000     & 0.018 & 0.054 & 0.000 & 0.017 & 0.068 & {0.000}\\
          
          & DeepSeek-V3 & 0.015 & 0.071 & 0.000 & \cellcolor{backblue!75}{0.019} & {0.060} & 0.000 & 0.017 & 0.066 & 0.000 \\
          
          & DeepSeek-R1 & 0.014 & \cellcolor{backblue!50}{0.081} & 0.000 & 0.015 & \cellcolor{backred!50}{0.068} & 0.000 & 0.015 & \cellcolor{backred!50}{0.075} & 0.000\\
          
          & Llama-3.1-70B & 0.014 & 0.061 & 0.000     & 0.015 & 0.046 & 0.000 & 0.015 & 0.054 & {0.000}\\
    \midrule
    \multirow{7}[2]{*}{\rotatebox{90}{Book}} 
    & GPT-4o-mini & 0.038 & 0.104 & \cellcolor{backblue!75}{0.004} & -     & -     & -  & 0.038 & 0.104 & \cellcolor{backblue!75}{0.004}\\
          & GPT-4o & 0.043 & 0.101 & 0.022 & -     & -     & - & 0.043 & 0.101 & 0.022\\
          & Gemini & \cellcolor{backred!50}{0.056} & \cellcolor{backred!50}{0.127} & 0.049 & -     & -     & - & \cellcolor{backred!50}{0.056} & \cellcolor{backred!50}{0.127} & 0.049\\
          & Claude &  0.018     &   0.072    &   0.012   & -     & -     & - & 0.018     &   0.072    &   0.012\\
          
          & DeepSeek-V3 & 0.020 & 0.081 & 0.005 & - & - & - & 0.020 & 0.081 & 0.005\\
          
          & DeepSeek-R1 & 0.030 & \cellcolor{backblue!75}{0.112} & 0.031 & -     & -     & - & 0.030 & \cellcolor{backblue!75}{0.112} & 0.031\\
          & Llama-3.1-70B & \cellcolor{backblue!75}{0.049} & 0.098 & \cellcolor{backred!50}{0.003} & -     & -     & - & \cellcolor{backblue!75}{0.049} & 0.098 & \cellcolor{backred!50}{0.003}\\
    \bottomrule
    \end{tabular}%
    }
  \label{tab:collaborative_performance}%
  \vspace{-1.5em}
\end{table}%

\noindent \textbf{\underline{Observation 8.} For different query types, Demographics-based Query exhibit lower Recall than Interest-based Query but similar Precision.}
Across all models, Recall for Demographics-based Query is generally lower than for Interest-based Query. This suggests that LLMs struggle to identify relevant recommendations when relying on user demographics. The difference in Recall can be attributed to the distinct ways in which these queries are constructed. Interest-based Query leverages clear interest, making it easier for models to generate relevant recommendations. In contrast, Demographics-based Query requires recommendation assistants to infer preferences from broader demographic and behavioral trends, which introduces more variability and ambiguity.

\begin{figure}[t]
\vspace{-1.5em}
    \centering
    \includegraphics[width=0.85\linewidth]{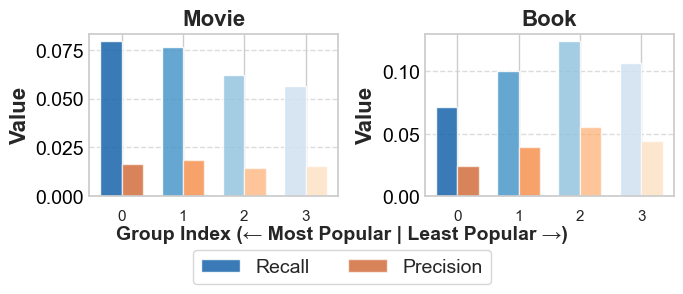}
    \vspace{-1.7em}
    \caption{Impact of Interest Popularity on Precision \& Recall}
    \label{fig:item-popularity}
    \vspace{-1.5em}
\end{figure}

\noindent \textbf{\underline{Observation 9.} For Interest-based Query, queries constructed from less prevalent interests tend to be more challenging.}
As shown in Figure~\ref{fig:item-popularity}, we categorized the queries into four groups based on their popularity and conducted experiments on the movie and book datasets.
It can be observed that interests that are shared by more users tend to yield higher Precision and Recall. This is likely because widely shared interests are more easily recognizable and interpretable by LLMs, leading to better performance.
In contrast, the book domain shows an opposite trend. This is due to the nature of book data and the evaluation process. Books often have multiple editions or publishers, and popular titles tend to have more variants. Therefore, the popular book (e.g., ``Dune'') can suffer lower metrics because LLMs confuse editions/publishers (e.g., ``Dune: 1965'' vs. ``2021 adaptation'').
% As a result, LLMs struggle to match the correct version to the ground truth.
For less popular books, fewer variants exist, making it easier for LLMs to provide exact matches, which leads to higher performance metrics.

\begin{figure}[h]
    \centering
    \vspace{-1.2em}
    \includegraphics[width=0.85\linewidth]{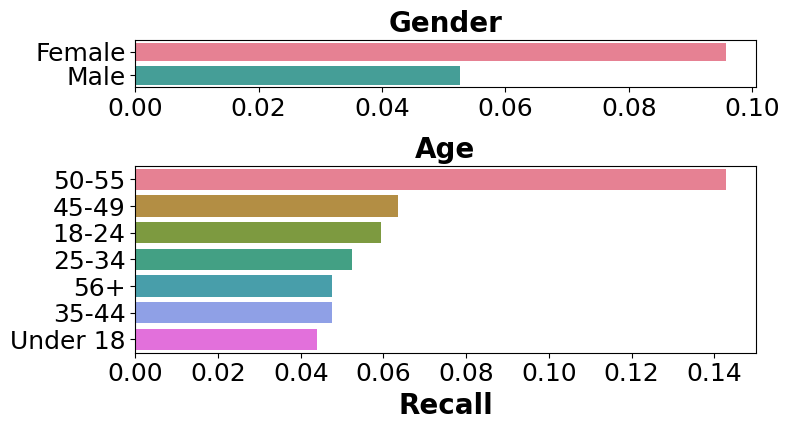}
    \vspace{-1.8em}
    \caption{Performance of queries with various demographics.}
    \label{fig:userbased_analysis}
    \vspace{-1em}
\end{figure}
\noindent \textbf{\underline{Observation 10.} For Demographics-based Query, the performances of LLMs show notable variations based on different user demographics.} 
For instance, as figure~\ref{fig:userbased_analysis} shows, we note that models exhibit higher accuracy for female users. It reflects more consistent preference patterns of female users (e.g., stable genre preferences like romantic comedies), which allows the model to make accurate recommendations.
From the perspective of age, LLMs perform better regarding Recall for people aged 50–55. This is likely because their preferences are more focused and less affected by rapidly changing popular culture compared to younger users, who tend to have more diverse preferences.
For users aged 56 and above, the model's effectiveness is not as strong as it is for the 50–55 age group, possibly because older users engage less frequently in online activities, such as rating movies. As a result, there is less data about this age group in the training set of LLMs, which makes it more challenging for LLMs to understand their preferences. 
% \sj{***This is not a good analysis. Consider more in-depth analyses like "other age people exhibit fragmented or underrepresented preferences."***}

\begin{table}[b]
  \centering
    \vspace{-1.2em}
  \caption{Fine-tuning on User Profile-based Query}
  \vspace{-1.5em}
  \resizebox{0.45\textwidth}{!}{
    \begin{tabular}{ccccc}
    \toprule
    \multicolumn{1}{l}{Domain} & \multicolumn{1}{c}{Model} & Fine-tuning & \multicolumn{1}{c}{Interest-based} & \multicolumn{1}{c}{Demographics-based} \\
    \midrule
    \multirow{6}[4]{*}{Movie} & \multirow{3}[2]{*}{Qwen-2.5-3B} & -     &  0.015     & 0.029 \\
          &       & SFT   &   \underline{0.287}    & 0.154  \\
          &       & RFT    &   0.143    &  \underline{0.228} \\
          &       & SFT+RFT    &  \textbf{0.289}     & \textbf{0.230}   \\
\cmidrule{2-5}          & \multirow{3}[2]{*}{Llama-3.2-3B} & -     &  0.014     & 0.019 \\
          &       & SFT   &   \underline{0.139}    & 0.102 \\
          &       & RFT    &    0.111      & \underline{0.122} \\
          &       & SFT+RFT    &  \textbf{0.144}     & \textbf{0.131}   \\
    \midrule
    \multirow{6}[4]{*}{Book} & \multirow{3}[2]{*}{Qwen-2.5-3B} & -     &  0.026     & - \\
          &       & SFT   &    \underline{0.268}   & - \\
          &       & RFT    &  0.108     & - \\
          &       & SFT+RFT    &  \textbf{0.270}     & -   \\
\cmidrule{2-5}          & \multirow{3}[2]{*}{Llama-3.2-3B} & -     &   0.022    & - \\
          &       & SFT   &   \underline{0.189}    & - \\
          &       & RFT    &   0.095    & -  \\
          &       & SFT+RFT    &  \textbf{0.193}     & -   \\
    \bottomrule
    \end{tabular}%
    }
    \vspace{-1.2em}
  \label{tab:collaborative-finetune}%
\end{table}%

\noindent  \textbf{\underline{Observation 11.} For User Profile-based Query, both fine-tuning methods lead to clear performance gains.} As shown in Table~\ref{tab:collaborative-finetune}, all fine-tuned models consistently outperform the base model. The SFT+RFT approach delivers the strongest overall performance, with SFT close behind. However, in certain cases—such as demographics-based queries—RFT outperforms SFT; for example, in the Qwen-2.5-3B model, RFT achieves a recall of 0.228 compared to SFT’s 0.154. We also observe that after fine-tuning, Qwen models generally show better results than the LLaMA models.

\vspace{-0.5em}
\section{Related Work}
Recent advances in LLM-based recommender systems~\cite{,wang2025knowledge,qu2025generative,qu2024tokenrec} have led to the development of various benchmarks to evaluate their capabilities~\cite{liu2023llmrec,jiang2024beyond,tan2025can,xue2025mmrc,zhang2023chatgpt}. 
Early efforts like LLMRec~\cite{liu2023llmrec} focus on traditional recommendation tasks (e.g., rating prediction, sequential recommendation), establishing baselines for LLMs against classical methods.
In addition to traditional recommendation tasks, recent studies have increasingly focused on evaluating the unique characteristics of LLM for recommendations. For instance, one recent study~\cite{jiang2024beyond} introduces a multi-dimensional framework that assesses not only utility and novelty but also LLM's specific behaviors like history length sensitivity (how recommendations perform with sparse user histories) and hallucination (generation of non-existent items). 
Similarly,  PerRecBench~\cite{tan2025can} isolates personalization accuracy by eliminating biases in users and item ratings, testing whether LLMs infer user preferences.

Meanwhile, with the rise of LLMs, varying benchmarks have been proposed to evaluate their capabilities as assistants. For example, Gaia~\cite{mialon2023gaia} measures LLM's capabilities as a general-purpose AI assistant. In addition, domain-specific frameworks like VoiceBench (for voice assistant)~\cite{deng2024mobile} and Mobile-Bench (for mobile assistant)~\cite{chen2024voicebench} assess the ability of LLM as the specific assistant.
However, despite progress in both recommendation and assistant evaluation, there remains a gap in assessing LLMs as personalized recommendation assistants.
To bridge this gap, we introduce \ourname, a novel benchmark designed to evaluate LLMs’ potential as personalized recommendation assistants.

\section{Conclusion}
In this paper, we introduce a new paradigm for next-generation recommender systems and propose a novel benchmark dataset, \ourname, to evaluate the potential of LLMs as personalized recommendation assistants. 
Our benchmark includes five specially designed query types that comprehensively assess LLM-based assistants' capabilities in handling complex user queries in real-world scenarios. We conduct extensive experiments on a diverse set of LLMs, including those with advanced reasoning capabilities. Through these experiments, we present 11 detailed observations that shed light on the strengths and limitations of LLMs in serving as personalized recommendation assistants. We also explore the impact of different fine-tuning strategies on enhancing recommendation performance. We hope our benchmark and findings lay a foundation for future research and development in this domain.

\begin{acks}
The research described in this paper has been partially supported by the General Research Funds from the Hong Kong Research Grants Council (project no. PolyU 15207322, 15200023, 15206024, and 15224524), internal research funds from Hong Kong Polytechnic University (project no. P0042693, P0048625, and P0051361).  
This work was supported by computational resources provided by The Centre for Large AI Models (CLAIM) of The Hong Kong Polytechnic University.
\end{acks}
\appendix
% \newpage
\vspace{1em}
\noindent\textbf{\LARGE APPENDIX}
% \vspace{-1em}
\section{Prompts for Model Testing}
\label{app:test_prompts}
The prompts we use for testing different LLMs are:

\begin{itemize}[leftmargin=*]
    \item w/o user-item interaction history: You are a movie recommender system. You are given a \underline{query} from the user. You need to recommend movies that satisfy the user's query. Please output the recommended movies directly; do not output other content.
    
    \item with user-item interaction history: You are a movie (/book) recommender system. You are given \underline{a user's watched movies} (/reading history) and a \underline{query} from the user. You need to recommend movies (/books) that satisfy the user's query and user preferences. Please output the recommended movies directly; do not output other content.
\end{itemize}

\section{Case Study}
\label{app:case study}
We present a case study to illustrate how LLMs handle complex user queries. In this example, the user seeks movies with the same cinematographer as two given films. The model DeepSeek-V3, lacking strong reasoning abilities, mistakenly lists movies by \textit{Scott Ambrozy}, leading to incorrect recommendations. In contrast, DeepSeek-R1 first verifies whether \textit{Scott Ambrozy} is the cinematographer, then correctly identifies \textit{Owen Roizman} as the actual cinematographer. It finally infers that the user intended to ask about \textit{Roizman}’s films, providing accurate recommendations.

\begin{figure}[t]
\footnotesize

\begin{cmt*}{Case Study}{}
\textbf{Query:} I recently enjoyed the \textbf{cinematography in Taps (1981) and Absence of Malice (1981)}, both of which showcased \textcolor{red}{Scott Ambrozy}'s incredible work. I'm eager to find more films that feature his cinematography style. Any recommendations?
\vspace{2pt}
\\
\textbf{Ground Truth: } 3 Days of the Condor (a.k.a. Three Days of the Condor) (1975), Network (1976), The Exorcist (1973), The Stepford Wives (1975) ; or "None"
\vspace{10pt}
\\
\textbf{Output (DeepSeek-V3): }\textit{The Chosen (1981), Four Friends (1981), Split Image (1982), Heart Like a Wheel (1983)} \redcross
\vspace{10pt}
\\
\textbf{Output (DeepSeek-R1): } 
\textit{The French Connection (1971), \textcolor{mygreen}{The Exorcist (1973)}, \textcolor{mygreen}{Network (1976)}, \textcolor{mygreen}{Three Days of the Condor (1975)}, The Electric Horseman (1979), Straight Time (1978), The Prince of the City (1981), Wyatt Earp (1994).}  

\smallskip % 使用 \smallskip 代替空行
\textit{Note: The films listed above feature the cinematography of \textit{Owen Roizman}, who shot \textit{Taps} and \textit{Absence of Malice}. Scott Ambrozy appears to be a confusion or typo in the query.}
\greentick

\textbf{Reasoning (DeepSeek-R1): }\textit{... Wait, first, who is Scott Ambrozy? I know some big cinematographers, but this name isn't ringing a bell. Maybe I should verify if there's a Scott Ambrozy in the industry. Let me think. Maybe the user misspelled the name? Looking up Taps on IMDb, yeah, Owen Roizman is the cinematographer for Taps. Absence of Malice also lists Owen Roizman. So Scott Ambrozy might be a confusion with another name. \textbf{So the user probably meant Owen Roizman.} So to answer correctly, \textbf{I should recommend other films by Owen Roizman}. Since the user wants movies with his cinematography style, I should list those. Roizman's notable works include...} 
\end{cmt*}
 \end{figure}

\newpage
\noindent\textbf{\Large Ethical Considerations}

Our work utilizes an LLM and public recommendation datasets for query generation. We have described the details of the generation process of our dataset in the main text of our paper. It is worth noting that all data in our dataset has been checked by human annotators to avoid harmful content. Therefore, our dataset does not pose any potential harm to human society.

\bibliographystyle{ACM-Reference-Format}
\balance
\bibliography{ref}


\end{document}